# Superior thermal conductivity of poly (ethylene oxide) for solid-state electrolytes: a molecular dynamics study

Han Meng[1,2], Xiaoxiang Yu[1,2], Hao Feng[1,2], Zhigang Xue[3,*], Nuo Yang[1,2,*]

[1] *State Key Laboratory of Coal Combustion, Huazhong University of Science and Technology, Wuhan 430074, P. R. China*

[2] *Nano Interface Center for Energy, School of Energy and Power Engineering, Huazhong University of Science and Technology, Wuhan 430074, P. R. China*

[3] *Key Laboratory for Material Chemistry of Energy Conversion and Storage, Ministry of Education, School of Chemistry and Chemical Engineering, Huazhong University of Science and Technology, Wuhan 430074, P. R. China.*

\* Corresponding E-mail: zgxue@hust.edu.cn (Z. X.); nuo@hust.edu.cn (N. Y.)

# Abstract

Solid-state lithium-ion batteries (SSLIBs) are considered to be the new generation of devices for energy storage due to better performance and safety. Poly (ethylene oxide) (PEO) based material becomes one of the best candidate of solid electrolytes, while its thermal conductivity is crucial to heat dissipation inside batteries. In this work, we study the thermal conductivity of PEO by molecular dynamics simulation. By enhancing the structure order, thermal conductivity of aligned crystalline PEO is obtained as high as 60 $Wm^{-1}K^{-1}$ at room temperature, which is two orders higher than the value (0.37 $Wm^{-1}K^{-1}$) of amorphous structure. Interestingly, thermal conductivity of ordered structure shows a significant stepwise negative temperature dependence, which is attributed to the temperature-induced morphology change. Our study offers useful insights into the fundamental mechanisms that govern the thermal conductivity of PEO but not hinder the ionic transport, which can be used for the thermal management and further optimization of high-performance SSLIBs.

# Introduction

Lithium-ion batteries (LIBs) have been widely applied in a diverse range, from portable devices to electric vehicles.[1] Recently, solid-state lithium-ion batteries (SSLIBs) with nonflammable solid electrolytes are becoming increasingly attractive, due to their better safety and higher performance compared with liquid-state ones.[2] However, thermal issue still remains and becomes increasingly prominent. As energy density increases and package volume reduces, the enormous heat generates inside SSLIBs. The poor thermal transport inside batteries will not only suppress heat dissipation but also increase temperature inhomogeneity and thermal stress inside batteries, which can significantly deteriorates cycle life and may trigger thermal runaway. With respect to this, thermal management of LIBs is very crucial to their safety and performance.[3-6]

Over the past decades, most of the researches have focused on the design, optimization and preparation of electrodes and electrolytes.[7] Only a few works have paid attention to the thermal management of LIBs. Some efforts aimed at modeling of external cooling technologies, such as forced air and liquid cooling, to lower the temperature.[8-11] As for the thermal transport inside batteries, researches mainly focus on liquid-state LIBs. Yang et al measured and identified that the separator is a major limiting factor for heat dissipation in liquid state LIBs, therefore a hierarchical polymer separator was prepared to obtain higher thermal conductivity.[12] A novel system that incorporates phase-change material was proposed to store and utilize the heat generated inside LIBs.[13] However, thermal transport inside SSLIBs is less investigated.

Different from cathodes and anodes, solid-state electrolytes (as shown in Figure 1(a)) are commonly formed by dissolving a salt in a solid host polymer matrix, which constitutes the frameworks of the composite materials. Among all the polymer materials, Poly (ethylene oxide) (PEO) is considered as a promising candidate of host polymer for high energy density solid electrolyte, since it have high energy density, good electrochemical stability, excellent compatibility with lithium salts, high safety, easy fabrication and low cost.[14] However, the thermal transport inside PEO based electrolyte is significantly hindered by the polymer matrix with low thermal conductivity. That is, there is a necessity to investigate and improve the thermal properties of PEO.

Meanwhile, it is worth noting that the way of enhancing thermal conductivity should not hinder ionic transport which is crucial to the performance of batteries. Previous studies suggest that ionic transport can be facilitated in crystalline PEO with ordered nano-channels, since the nano-channels results an increase in ionic conductivity up to 30 times higher than traditional complex/amorphous structure.[15, 16] On the other side, an order structure in atomic scale is also benefit for the thermal transport, due to the decrease of phonon scattering. Therefore, a crystalline structure with directionally aligned PEO chains is a good candidate to achieve high thermal conductivity for solid electrolyte.

In this letter, we investigated numerically the thermal conductivities of PEO in the preliminary stage. We firstly constructed an amorphous PEO (APEO) structure and obtained a poor value of thermal conductivity. Then, we construct an ordered crystalline PEO (OCPEO) structure by aligning polymer chains to achieve better heat transport, and studied the thermal conductivity. Lastly, we performed morphology analysis on PEO chains to interpret the temperature dependence of thermal conductivity by quantifying the radius of gyration, volumetric thermal expansion and radial distribution function.

## Method and model

The equilibrium molecular dynamics (EMD) simulation method (Green-Kubo method), which has been widely used to calculate thermal transport properties.[17-20] Green–Kubo formula[21, 22] is a result of the linear response theory and the fluctuation dissipation theorem, which relates the heat flux autocorrelation to the thermal conductivity. The heat current is defined as

$$\vec{J} = \frac{1}{V}\left[\sum_i e_i \vec{v}_i + \frac{1}{2}\sum_i \vec{r}_{ij}(\vec{F}_{ij} \cdot \vec{v}_i)\right], \tag{1}$$

and the thermal conductivity is derived from the Green-Kubo equation as

$$\kappa = \frac{V}{3k_B T^2} \int_0^{\tau_0} \langle \vec{J}(0) \cdot \vec{J}(\tau) \rangle \, d\tau, \tag{2}$$

where $k_B$ is the Boltzmann constant, $V$ is the system volume, $T$ is the temperature, $\tau$ is the correlation time, $\tau_0$ is the integral upper limit of heat current auto-correlation function (HCACF), and the angular bracket denotes an ensemble average. Generally, the temperature in MD simulation is calculated by the formula

$$\langle E \rangle = \sum_{i=1}^N \frac{1}{2} m_i v_i^2 = \frac{3}{2} N k_B T_{MD}, \tag{3}$$

where $E$ is total kinetic energy of the group of atoms, and $N$ is number of total atoms.

The unit cell of PEO consist of two singly-bonded carbon atoms alternately connected by an oxygen atom, as shown in Figure 1(b). All EMD simulations are performed by the large-scale atomic/molecular massively parallel simulator (LAMMPS) package.[23] The polymer consistent force-field (PCFF)[24, 25] is used to describe interatomic interactions, which includes anharmonic bonding terms and is intended for applications in polymers and organic materials. Periodic boundary conditions are applied in all three dimensions. And the velocity Verlet algorithm [26] is employed to integrate equations of motion. 0.25 fs and 10 Å are chosen as time step and cutoff distance for the Lennard-Jones interaction respectively. In addition, 6 independent simulations with different initial conditions are conducted to get better average.

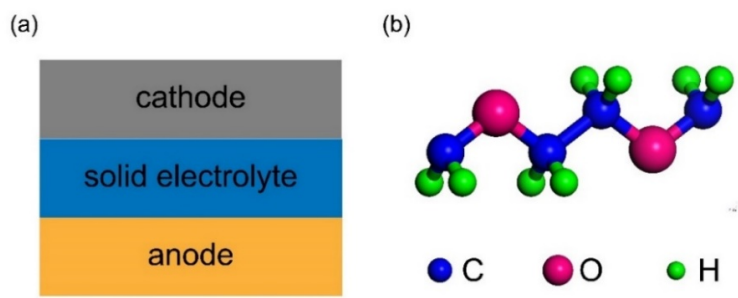

**Figure 1.** (a) The typical structure of solid-state lithium batteries; (b) The unit cell of PEO, consisting of carbon (gray), oxygen (red) and hydrogen (white) atoms.

The APEO is made from 40 single PEO chains with each chain containing 100 carbon atoms and 50 oxygen atoms. The preparation procedure of amorphous structure is shown in Figure 2. Initially, a single extended PEO chain is simulated and equilibrated at 300 K for 1 ns to form a compacted particle. Then 40 of these particles are randomly packed into a supercell. After minimization, a constant number of atoms, pressure and temperature (NPT) ensemble is used to increase the system temperature from 300 K to 600 K by a rate of 50 K/ns, and a 12 ns NPT ensemble at 600 K is used to generate PEO melt with fully relaxed and amorphous structure. The obtained structure is then quenched to different target temperatures, and NPT ensemble runs for 1 ns are used to further equilibrate the structures at the quenched temperatures. After the stable structures are obtained, the NVE ensemble runs for 1 ns are used to record heat flux and calculate thermal conductivity. Due to the isotropic structure of APEO, thermal conductivity along all three directions are used to obtain an average value.

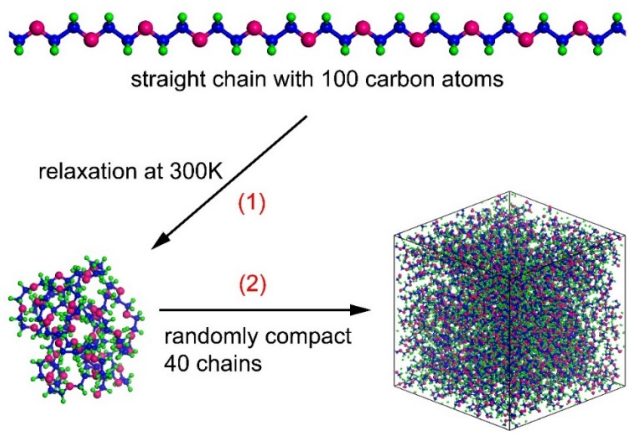

**Figure 2.** Illustration of initial APEO structure preparation.

The initial crystalline structure is constructed by aligning straight PEO chains (as shown in Figure 4(a)). The systems are firstly simulated in NPT ensembles at target temperatures and 1 atm for 100 ps to obtain the optimized structures and simulation cell sizes, then follow by NVE ensembles for 100 ps before collecting heat flux data in NVE ensembles for 3 ns. Noting that the thermal conductivity of OCPEO is highly anisotropic, namely the radial-direction (y and z) values is much lower than the corresponding chain-direction (x) value, so we only focus on the chain-direction value.

## Results and discussion

To ensure the amorphous structures are well equilibrated, the radius of gyration is calculated for the characterization of system morphology. As shown in Figure 3(a), the reached oscillating convergence indicates the equilibrium of the amorphous structure. After the structure are well equilibrated, we calculated the thermal conductivity of APEO at 300 K and the results are depicted in Figure 3(b) as an illustration of the EMD method. The typical normalized heat current auto-correlation function (HCACF) fluctuates dramatically for 0.5 ps, which means the strong reflection of heat current, and then it decays to zero rapidly within 2 ps for the sake of strong phonon scatterings in APEO, thus giving rise to the quick convergence of integral, *i.e.* thermal conductivity value.

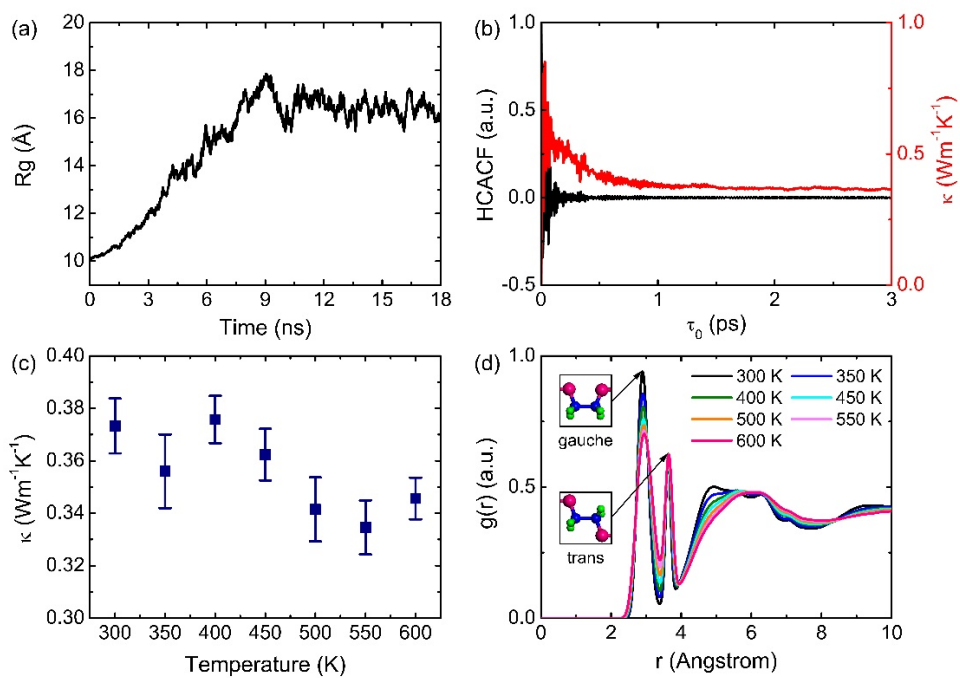

**Figure 3.** (a) Average radius of gyration of polymer chains during heat treatment; (b) Normalized HCACF (blue line) and integral thermal conductivity (black line); (c) Thermal conductivity of APEO as a function of temperature; (d) Radial distribution function of APEO at different temperature.

In practical terms, temperature of batteries raises with the increase of charging and discharging time. Therefore, we study the thermal conductivity of APEO and its temperature dependence. As shown in Figure 3(c), the value of thermal conductivity is obtained as 0.37±0.01 Wm$^{-1}$K$^{-1}$ at room temperature, which is on the same order of magnitude as bulk epoxy[27], cross-linked PE[28] and hydrogel[29]. Thermal conductivity of APEO shows a weak negative dependence on temperature. As given in Figure 3(d), we also calculate the three dimensional radial distribution function (RDF) of oxygen atoms at different temperature, the insets show the *trans* and *gauche* conformation. With temperature increases, the peak around 3 Å (*gauche* conformation) reduces and the peak around 3.8 Å (*trans* conformation) stay unchanged, but the trough around 3.5 Å increases, which indicates more chain segments maintain between *gauche* and *trans* conformation. So the weak negative temperature dependent thermal conductivity is ascribed to the increasing phonon scattering induced by intermediate chain conformation at high temperature. As a result of the severe phonon scattering induced by disordered structure, such a low thermal conductivity is bad for the heat dissipation in SSLIBs and it need to be further improved.

The enhancement of structure order can improve thermal conductivity, because phonon scatterings can be reduced. Therefore, we construct a crystalline structure by aligning PEO chains in a triclinic lattice, as shown in Figure 4(a). When using Green-Kubo formula to calculate thermal conductivity, the finite size effect would arise if the simulation cell is not large enough. Before studying the thermal conductivity by EMD simulation, the simulation cell size is checked to overcome the finite size effect and obtain a converged value of thermal conductivity. As shown in Figure 4(b), we calculated the thermal conductivity of OCPEO at 300 K with different chain length (CL) and cross section area (CSA). It is found that the thermal conductivity along the chain direction is independent on CSA but dependent on CL, which converges at about 20 nm. Therefore, all subsequent simulations use a system with CL of 20 nm and CSA of 5 $nm^2$.

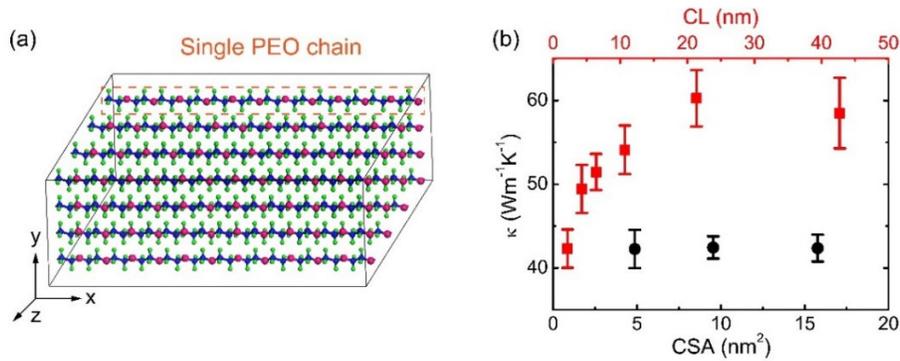

**Figure 4.** (a) Illustration of a OCPEO structure; (b) Thermal conductivity of OCPEO versus chain length (red squares, where the CSA is 5 nm$^2$) and cross section area (black dots, where the CL is 2 nm) at 300 K.

And then, the temperature dependence of thermal conductivity of OCPEO is investigated, which is one of the main results. As shown in Figure 5(a), the thermal conductivities of OCPEO at different temperature are calculated, and the value of 60±3 Wm$^{-1}$K$^{-1}$ is obtained at room temperature, which is two orders of magnitude higher than that of APEO. When compare with PE composites, the value is smaller than that of paved crosswise polyethylene laminate (181 Wm$^{-1}$K$^{-1}$) [30], PE nanofibers (104 Wm$^{-1}$K$^{-1}$) [31], aligned carbon nanotube-PE composites (99.5 Wm$^{-1}$K$^{-1}$) [32] and Br-doped graphene fiber (95.8 Wm$^{-1}$K$^{-1}$) [33], but larger than PE nanowire arrays (21.1 Wm$^{-1}$K$^{-1}$) [34] at room temperature. With temperature increases from 300 K to 600 K, the thermal conductivity is sharply reduced by two order of magnitude. Noting that generally in bulk structure, the thermal conductivity decrease as ~T$^{-1}$ due to the enhancement of Umklapp phonon-phonon scattering at high temperature. Differently, there is a further decrease of thermal conductivity, and it shows a stepwise trend separated by two abrupt drops around 490 K and 550

K, where the thermal conductivity decreases almost an order of magnitude. The system structures are shown in Figure 5(b), which exhibit three distinctive morphologies in different temperature regions. Such a temperature dependence of thermal conductivity is believed to be related to the temperature-induced morphology change[35], which will be analyzed in following. And the similar stepwise trend of thermal conductivity was also found in polyethylene and nylon.[36, 37]

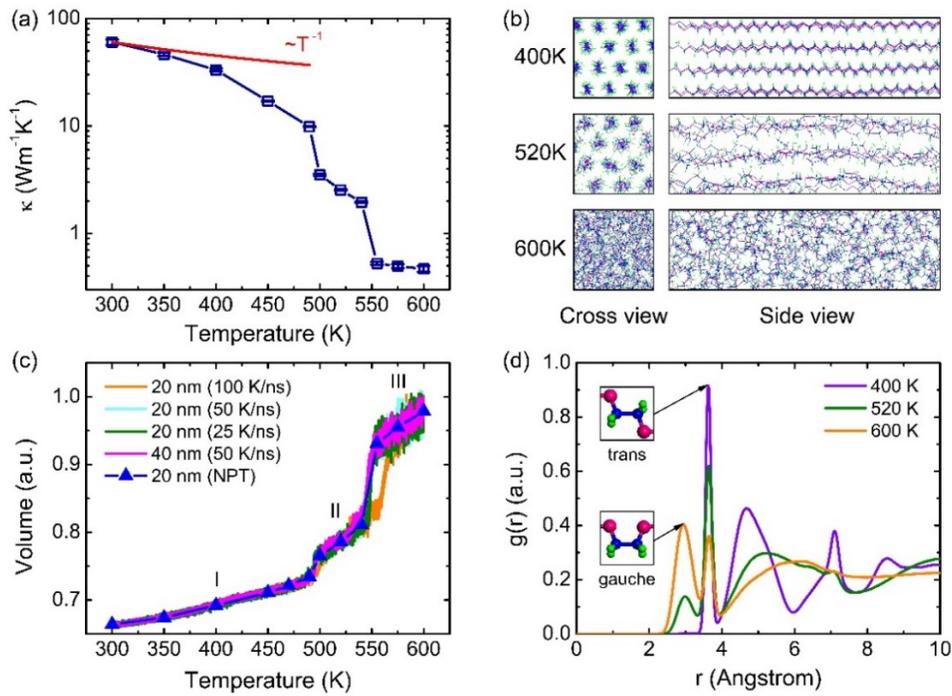

**Figure 5.** (a) Thermal conductivity of OCPEO (blue squares) and $\sim T^{-1}$ fitting (red line) as a function of temperature; (b) Cross and side views of OCPEO at 400 K, 520 K, and 600 K; (c) Normalized simulation volume of OCPEO as a function of temperature; (d) Radial distribution function of OCPEO at 400 K, 520 K, and 600 K.

To study the temperature dependent morphology change, the system volume is recorded during OCPEO is heated from 300 K to 600 K with a heating rate of 50 K/ns at a constant pressure of 1 atm. The volume profile remains almost unchanged when heating rate is half or simulation cell size in the chain direction (x direction) is doubled, which indicates that heating process is slow enough and simulation size is large enough. As shown in Figure 5(c), the volume increase with the temperature linearly in three regions, which are divided by two obvious jumps around 490 K and 550 K. The system volume after structure optimization are also recorded at different temperature, and the volume jumps are reproduced by these NPT ensemble around 490K and 550K, which are consistent with the two temperature values in volume profile, respectively. The structure order are quite different in three temperature regions. In region I (300 K-490 K), the atoms vibrate around the equilibrium positions, both inter-chain and along-chain crystal structure are well maintained. In region II (490 K-550 K), all atoms can vibrate in a larger range than in region I and chain segments can move in a small range. The along-chain lattice order is destroyed, however the inter-chain positions is not destroyed and still displays lattice order, which is observed along the cross-sectional direction (y-z plane). In this region, chains are still aligned without tangling. In region III (>550 K), the system is melted and the chains are tangled and the structure is change to an amorphous phase.

To better show the morphology change, we further calculated the three dimensional RDF of oxygen atoms, which is used to quantify lattice orders at different temperatures. As shown in Figure 5(d), the RDF shows that lattice has a significant change from 400 K to 520 K. First, the peak around 5 Å shifting to the right, suggesting increase in the inter-chain distance. This creates more space for individual chains in the crystal and thus allows easier segment translation and rotation, which destroy the along-chain lattice order as demonstrated below. Second, the new peak generating around 3 Å and sharp peak reducing around 3.8 Å, which correspond to the lattice sites of oxygen atoms along PEO chains, indicating the transformation of conformation from *trans* to *gauche* in chain segments. Third, the peaks around 7 Å and 8.5 Å are significantly flattened, indicating the breakdown of certain lattice order. In general, the significant differences among RDF at different temperature demonstrates that lattice order of PEO crystal is changed. Such disorder can significantly scatter thermal transporting phonons along the chain, and thus reduces thermal conductivity. Therefore, the stepwise decrease in thermal conductivity can be attributed to morphology-induced phonon scattering as well as anharmonic phonon scattering, which increases with temperature. In practical, it is necessary to apply OCPEO below 490 K to maintain the high thermal transport performance.

## Conclusion

In general, we systematically study the thermal conductivity of PEO by performing EMD simulations. We calculate the thermal conductivity for APEO, the value of 0.37±0.01 Wm$^{-1}$K$^{-1}$ is obtained at room temperature, and it shows a weak dependence on temperature. For the purpose of applying PEO as solid electrolyte for SSLIBs, thermal conductivity needs to be further improved. We construct an OCPEO in order to improve thermal conductivity, the value of 60±3 Wm$^{-1}$K$^{-1}$ is obtained at room temperature, which is two orders of magnitude higher than that of APEO. With the increase of temperature, thermal conductivity of OCPEO shows a stepwise decrease trend separated by two abrupt drops. By recording volume during heating and calculating the RDF at different temperature, we demonstrate that the stepwise thermal conductivity is attribute to the temperature-induced morphology change. Our results define a direction in the search for high thermal conductive ionically conducting PEO, which emphasizes structure order as important features. Insights obtained from this study can be used for the thermal management and further optimization for high-performance SSLIBs. Besides, the influence of lithium salt on thermal conductivity of PEO need to be studied in the future work.

# Conflicts of interest

There are no conflicts of interest to declare.

# Acknowledgements

This work is financially supported by National Natural Science Foundation of China (No. 51576076, No. 51711540031 and No. 51622303), Natural Science Foundation of Hubei Province (2017CFA046) and Fundamental Research Funds for the Central Universities (2016YXZD006). We are grateful to Meng An, Dengke Ma and Xiao Wan for useful discussions. The authors thank the National Supercomputing Center in Tianjin (NSCC-TJ) and China Scientific Computing Grid (ScGrid) for providing assistance in computations.